\documentclass[aps, 11pt, preprint, prc, amsmath, amssymb, nofootinbib, superscriptaddress]{revtex4}

\usepackage[usenames]{color}
\usepackage{graphicx}
\usepackage{amsmath}
\usepackage{amsfonts}
\usepackage{amssymb}
\usepackage{mathrsfs}
\usepackage{bm}
\usepackage{verbatim}
\usepackage{cancel}
\usepackage{CJK}
\usepackage{comment}
\usepackage[normalem]{ulem}
\usepackage[dvipsnames]{xcolor}
\usepackage{diagbox}

\newcommand{\chn}[3]{{{}^{#1}\!{#2}_{#3}}}
\newcommand{\cs}[2]{\chn{#1}{S}{#2}}
\newcommand{\cp}[2]{\chn{#1}{P}{#2}}

\newcommand{\NNLO}{{N$^2$LO}}

\newcommand{\Eref}{\mathcal{E}_r}
\graphicspath{{./figs/3p0_pert/}{./figs/1s0/}}

\begin{document}


\title{Trapped two-nucleon system in energy-dependent effective field theory}

\author{Chenbo Li}

\author{Jiexin Yu}

\author{Rui Peng}

\author{Songlin Lyu}
\email{songlin@scu.edu.cn}

\author{Bingwei Long}
\email{bingwei@scu.edu.cn}
\affiliation{College of Physics, Sichuan University, Chengdu, Sichuan 610065, China}

\date{October 7, 2021}

\begin{abstract}
We discuss how to connect the energy levels of two-particle systems trapped by a harmonic-oscillator force to scattering amplitudes, with nucleon-nucleon scattering phase shifts in uncoupled channels as the application. At the center of the proposed framework is the energy-dependent effective field theory that aims to expand observables in a neighborhood around each reference energy, often taken to be one of the energy levels. We also investigate how to disentangle the trapping force at short distances and the intrinsic interaction between the particles.
\end{abstract}

\maketitle


\section{Introduction\label{sec:intro}}

Many-body methods for studying nuclear structure have evolved to the point where properties of relatively tightly bound nuclei can be calculated with microscopic nuclear forces~\cite{Navratil:2007we, Epelbaum:2008ga, Ekstrom:2013kea, Barrett:2013nh, Lynn:2014zia, Ekstrom:2015rta, Piarulli:2017dwd, Hammer:2019poc}. But ab initio descriptions of nuclear reactions appear to be more difficult because of the larger size of reacting systems~\cite{Navratil:2016ycn, Yang:2016brl, Johnson:2019sps, Mazur:2020hdc, Ma:2020roi}. In a recent research program, efforts have been made to calculate scattering amplitudes, using as inputs the energy levels of many-nucleon systems trapped by an artificial field, especially a harmonic-oscillator (HO) potential~\cite{Rotureau:2010uz, Rotureau:2011vf, Luu:2010hw, Zhang:2019cai, Zhang:2020rhz, Guo:2021uig}. An indispensable ingredient in this approach is a model-independent formalism to connect the energy eigenvalue of the trapped states to the scattering amplitude of the two particles, the so-called quantization condition (QC). We propose in the paper an energy-dependent effective field theory (EDT) to achieve this goal, by expanding observables in a kinematic window centered around a reference center-of-mass (CM) energy. This reference energy can be most conveniently chosen to be one of the energy eigenvalues of the trapped system.

We illustrate the framework by studying elastic scattering of two particles that are often composite from the point of view of the underlying interaction applied in ab initio calculations. The EDT formalism developed here describes their interactions with contact operators; therefore, it appears to have the usual structure of cluster/halo effective field theory (EFT)~\cite{Hammer:2017tjm, Hammer:2019poc}. However, the EDT coupling constants depend on the reference CM energy $\Eref$ that anchors a particular kinematic neighborhood. The EDT can therefore be thought of as a collection of ``member'' EFTs that relay each other's validity region, and the full set of them as a whole covers a sufficiently large kinematic domain. Once a reference energy is chosen, the corresponding member EFT is in charge of the kinematic configurations that amount to small residual momenta around $\Eref$, forming expansion of observables in powers of $(E - \Eref)$, where $E$ is the CM energy of the said two particles. In this sense, the coupling constants of the EDT are still referred to in the paper as low-energy constants (LECs).

When applied to the trapped particles, the energy of an eigenstate under consideration can be designated as the reference energy $\Eref$. One collects inputs for the EDT from the eigenstate, provided by ab initio calculations, in order to determine the LECs at this reference energy. The idea of expanding around a reference energy enables us to make use of the contact EFT toolkit even away from the threshold, where cluster/halo EFT has been traditionally designed to function. We show that at leading order (LO) the formula first constructed in Ref.~\cite{Busch98}, referred to as the BERW formula in the paper, is established. More importantly, one can calculate reliably scattering amplitudes around $\Eref$ by investigating subleading corrections.

Among the previous studies on HO-trapped systems, EFT frameworks were frequently employed. In some works, the underlying interaction had a form of EFT and was at the center of investigation. So no effort was made to improve the BERW formula itself~\cite{Stetcu:2010xq, Rotureau:2010uz, Rotureau:2011vf}. In others, an EFT was used to solve simultaneously the two-body problems in the HO trap and free space~\cite{Zhang:2019cai, Zhang:2020rhz} so as to correlate the spectrum and elastic scattering amplitude, just like we will do in the paper. But the EFT expansion in Refs.~\cite{Zhang:2019cai, Zhang:2020rhz} focuses on small momenta near the threshold, conforming with the conventional wisdom of contact EFTs~\cite{vanKolck:1998bw, Kaplan:1998tg, Phillips:1997xu}.

It should be mentioned that the idea of building EFTs for kinematic configurations away from threshold has been implemented before in various applications. For instance, one can organize calculations in resonance and threshold regions separately before somehow combining them~\cite{Pascalutsa:2002pi, Long:2009wq, Long:2011rt}. An expansion of scattering amplitude can be constructed close to its zeros even though they are above the threshold~\cite{Lutz:1999yr}. In many-fermion systems, an EFT can be developed for small momentum modes near the filled Fermi surface~\cite{Polchinski:1992ed}.

Since the HO potential does not vanish within the range of intrinsic interactions for any finite value of HO frequency $\omega$, an additional price must be paid to disentangle the trapping and intrinsic forces. This was addressed in Ref.~\cite{Zhang:2019cai} by making the LECs of threshold EFT as a function of $\omega^2$. We approach the issue with the EDT framework, extrapolating scattering observables, rather than the LECs, for $\omega^2 \to 0$.

The EDT formalism is explained in Sec.~\ref{sec:framework} with the $\cs{1}{0}$ $NN$ scattering as the first example, followed by extension to uncoupled $P$ waves in Sec.~\ref{sec:3p0}. Section~\ref{sec:omega} demonstrates how to remove systematic errors due to finite $\omega$ by extracting the continuum limit of the $\cs{1}{0}$ scattering length. Some discussions and the conclusions are offered in Sec.~\ref{sec:conclusions}.

\section{Energy-dependent effective field theory\label{sec:framework}}

Among others, there are four important length scales at play inside a trap: the wavelength of the particles $\lambda$, the wideness of the trap $b$, the range of the intrinsic interaction $r_I$, and the size of the particles $d$. In this investigation, we assume that $d$ is much smaller than both $\lambda$ and $b$. The HO potential can be expressed in terms of reduced mass $\mu$ and interparticle distance $r$,
\begin{equation}
    V^\text{HO}(r) =   \frac{1}{2 }  \mu \omega^2 r^2 \, ,
\end{equation}
and the oscillator length is frequently used to characterize its wideness:
\begin{equation}
  b \equiv \left(\mu \omega\right)^{-\frac{1}{2}} \, .
\end{equation}
Connecting discrete energy spectra of artificially trapped particles to their scattering amplitudes finds many applications in various fields of physics~\cite{Luscher:1990ux, Elhatisari:2016hby}. Model-independent extraction of scattering information is most conveniently done when $b$ and $r_I$ are well separated: $r_I \ll b$. Therefore, there must be a noninteracting region where the intrinsic interaction at least nearly vanishes and the confining force of the trap is sufficiently weak. This allows one to construct infinite-volume scattering wave functions by matching their asymptotic form, with an undetermined phase shift, to the trapped ones in the noninteracting region.

It is perhaps easiest to construct the asymptotic wave functions with auxiliary potentials with contact operators~\cite{Zhang:2019cai, Guo:2020kph, Guo:2021uig}. As pointed out in Ref.~\cite{Guo:2021uig}, it matters little whether the auxiliary potential resembles the underlying, realistic interactions for distances shorter than $r_I$. More importantly, the contact parameters are not really ``constant''; it is absolutely fine for them to vary from one energy eigenvalue to another. We now build upon this idea a more systematic EFT framework that will include effective operators to describe small momentum fluctuations around a reference energy.

We use the two-nucleon system to illustrate the framework. More specifically, we consider the uncoupled channel $\cs{1}{0}$ of $NN$ scattering so that we will not be distracted by complications such as coupled-channel dynamics. As for the underlying nucleon-nucleon interaction, we choose the leading-order chiral force: one-pion exchange plus a constant contact term, regularized by a Gaussian function with $\Lambda = 400$ MeV, referred to below as $V^\text{LO}_{\chi}$. We note that, in this proof-of-principle exercise, any nucleon-nucleon potentials are acceptable.

We now turn to the EDT interaction that will underpin both scattering and trapped states. The $\cs{1}{0}$-projected EDT potential takes the usual form found in the plethora of contact EFT literature (for instance, see Refs.~\cite{vanKolck:1998bw, Kaplan:1998tg, Phillips:1997xu}), but with the LECs labeled by reference energies $\Eref$:
\begin{equation}
    V_\cs{1}{0} (p', p; \Eref) = C(\Eref) + \frac{1}{2}D(\Eref)(p'^2 + p^2) + \frac{1}{2}E(\Eref)p'^2\,p^2 + \cdots   \, , \label{eqn:VEDT}
\end{equation}
where $p$ ($p'$) is the incoming (outgoing) relative momentum. To calculate observables like scattering amplitudes or energy levels in the trap, the UV part of the EFT potential needs to be regularized
\begin{equation}
  V_\cs{1}{0}^{\Lambda} (p', p; \Eref) = f_R\left(\frac{{p'}^2}{\Lambda^2}\right) \, V_\cs{1}{0} (p', p; \Eref)\, f_R\left(\frac{{p}^2}{\Lambda^2}\right) \, .
  \label{eqn:regV}
\end{equation}
We use a Gaussian regulator in the numerical calculations carried out in the paper:
\begin{equation}
     f_R\left(x\right) = e^{-x^2} \, .
\end{equation}
The $\cs{1}{0}$ $T$-matrix is generated by the partial-wave Lippmann-Schwinger equation:
\begin{equation}
  T(p', p; E) = V(p', p) + \frac{1}{2\pi^2} \int_0^\infty dq q^2 V (p' , q) \frac{T(q, p; E)}{E - q^2/m_N + i0} \, ,
\end{equation}
which is sometimes written symbolically as
\begin{equation}
    T = V + V G_0(E) T \, ,
\end{equation}
where $G_0$ is the free-particle Green's function. The $T$-matrix is related to the phase shift by
\begin{equation}
    T = -\frac{4\pi}{m_N} \frac{1}{k \cot \delta -ik} \, ,
\end{equation}
where $E = k^2/m_N$.

The LO $T$-matrix is expected to be generated by the $C$ term alone. When calculating subleading corrections, one often finds it elucidating to treat higher-order operators in perturbation theory~\cite{vanKolck:2020llt}. The $T$-matrix at each order will calculated as follows:
\begin{align}
    T^{(0)} & = V_\cs{1}{0}^{(0)} + V_\cs{1}{0}^{(0)} G_0 T^{(0)}\, ,\\
    T^{(1)} & = (1 + T^{(0)} G_0) V_\cs{1}{0}^{(1)} (1 + G_0 T^{(0)}) \, , \\
    T^{(2)} & =  (1 + T^{(0)} G_0) V_\cs{1}{0}^{(2)} (1 + G_0 T^{(0)}) \notag \\
    & \quad + (1 + T^{(0)} G_0)V_\cs{1}{0}^{(1)} (G_0 + G_0 T^{(0)} G_0 ) V_\cs{1}{0}^{(1)} (1 + G_0 T^{(0)})\, ,
\end{align}
where
\begin{align}
    V_\cs{1}{0}^{(0)} &= C^{(0)} \label{eqn:V1s0LO}\, ,\\
    V_\cs{1}{0}^{(1)} &= C^{(1)} + \frac{1}{2}D^{(0)} (p'^2 + p^2) \, , \\
    V_\cs{1}{0}^{(2)} &= C^{(2)} + \frac{1}{2}D^{(1)} (p'^2 + p^2) + \frac{1}{2}E^{(0)} p'^2\,p^2  \, .
\end{align}
We have formally expand coupling constants at each order of EFT expansion, e.g.,
\begin{equation}
    C(\Eref) = C^{(0)}(\Eref) + C^{(1)}(\Eref) + C^{(2)}(\Eref) + \cdots \, , \label{eqn:ExpC}
\end{equation}
which does not, however, introduce more free parameters to the EFT~\cite{Long:2007vp}.

According to the power counting above, we show the first two orders explicitly. The LO $T$-matrix is found to be
\begin{equation}
T^{(0)}(E) = -\frac{4\pi}{m_N} \frac{1}{\alpha_0 (\Eref) - ik} \, ,
\end{equation}
where $E = k^2/m_N$ is the CM energy, and
\begin{equation}
\alpha_0(\Eref) = -\frac{4\pi}{m_N} \left[ C_0^{-1}(\Eref) + \theta_1 m_N \Lambda \right]\, , \label{eqn:alpha0}
\end{equation}
where $\theta_n$ depends on the regularization scheme
\begin{equation}
   \frac{1}{2\pi^2} \int \mathrm{d} q q^{n-1} f_R^2\left(\frac{q^2}{\Lambda^2}\right)  \equiv \theta_n \Lambda^n \, .
\end{equation}
We have discarded the terms vanishing for $\Lambda \to \infty$. The NLO correction is given by
\begin{equation}
    T^{(1)} = -\frac{4\pi}{m_N} \frac{-\alpha_1(\Eref)}{\left[\alpha_0 (\Eref) - ik\right]^2} (E - \Eref) \, ,
\end{equation}
where
\begin{equation}
    \alpha_1(\Eref) \equiv \frac{1}{4\pi} \frac{D^{(0)}}{{C^{(0)}}^{2}}\,, \qquad \Eref \equiv m_N \left(\theta_3 m_N C^{(0)}  \Lambda^3 - \frac{C^{(1)}}{D^{(0)}} \right) \, .
\end{equation}
The conversion of these $T$-matrices to the phase shifts must respect perturbative unitarity of the $S$-matrix so that its breaking is always in higher order by the same power counting~\cite{Long:2011xw, Wu:2018lai}. Therefore, we can rewrite the sum of $T^{(0)}$ and $T^{(1)}$ by adding necessary terms at {\NNLO} and beyond:
\begin{equation}
    T^{(0+1)} = -\frac{4\pi}{m_N} \frac{1}{\alpha_0(\Eref) + \alpha_1(\Eref) (E - \Eref) - ik }\, , \label{eqn:T1s0_exp}
\end{equation}
which resembles the effective range expansion
\begin{equation}
    k \cot \delta = \alpha_0(\Eref) + \alpha_1(\Eref) (E - \Eref) + \alpha_2(\Eref) (E - \Eref)^2 + \cdots \label{eqn:kcotExp}
\end{equation}

The energy levels of the trapped particles are generated by the total Hamiltonian of the trapping force and $NN$ interactions. The underlying interaction $V^\text{LO}_\chi$ is first put in the trap to generate energy eigenvalues that will later be used as  ``data.'' In order to determine the LECs of $V_\cs{1}{0}^\Lambda$, the energy spectrum of the following Hamiltonian must match the data, at least partially:
\begin{equation}
    H_\omega = H_0 + V^\text{HO} + V_\cs{1}{0}^\Lambda\, .
\end{equation}
To calculate subleading corrections to the eigenvalues, we treat higher-order terms as perturbations, in a fashion similar to scattering amplitudes.

The previous effort to demonstrate renormalizability of the EDT potential in $\cs{1}{0}$ scattering now pays off, for $b$ as an infrared scale will not change the ultraviolet behavior of EDT as long as $\Lambda \gg p \gtrsim b^{-1} = \sqrt{m_N \omega}/2$, where $p$ is the average momentum of the trapped state. So, we expect the energy eigenvalues $\mathcal{E}_n$ generated by $H_\omega$ are independent of $\Lambda$ at the limit $\Lambda \to \infty$. In the following, we will, as a convention, use large enough cutoff values in the EDT calculations to ensure that the results reach the limit $\Lambda \to \infty$.

\begin{figure}[tb]
  \centering
  \includegraphics[scale=0.45]{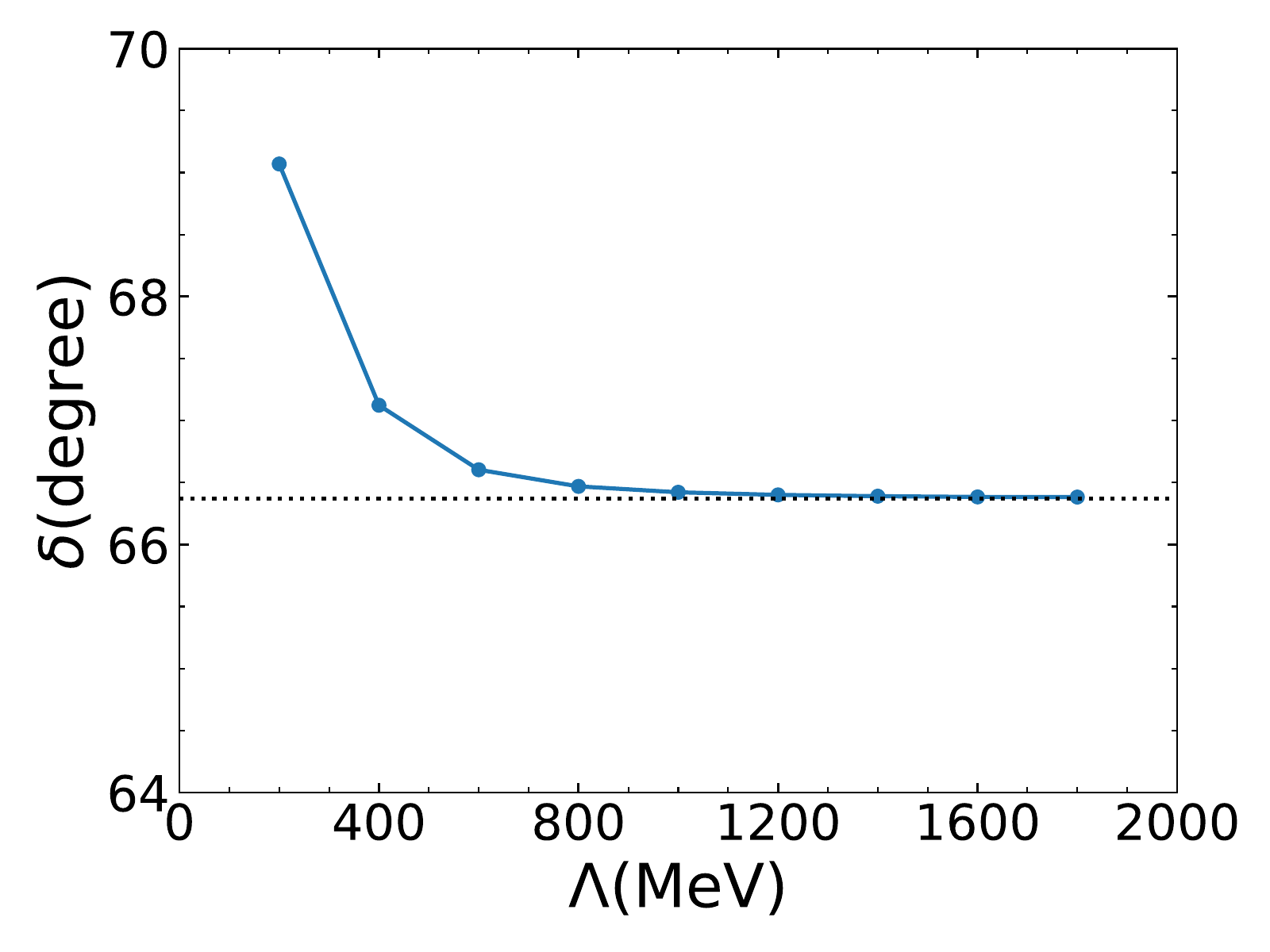}
  \caption{The LO $\cs{1}{0}$ phase shifts (the circles) approach the BERW value (the dotted line) for large cutoff values; the ground state for $\omega = 10$ MeV is used as the input. See the text for more explanation.}
  \label{fig_busch_lo_1s0}
\end{figure}

At LO, the EDT potential has only one operator with no explicit momentum dependence. This is precisely the auxiliary potential used in Ref.~\cite{Guo:2021uig}, with which the BERW formula was reproduced at the limit $\Lambda \to \infty$:
\begin{equation}
p^{2l+1} \cot \delta_l(\mathcal{E})=(-1)^{l+1}(4 \mu \omega)^{l+ \frac{1}{2}} \frac{\Gamma(\frac{3}{4}+\frac{l}{2}-\frac{\mathcal{E}}{2\omega})}{\Gamma(\frac{1}{4}-\frac{l}{2}-\frac{\mathcal{E}}{2\omega})}\, , \label{eqn:BERW}
\end{equation}
where $\mathcal{E}$ is an energy eigenvalue and $l$ is the orbital angular momentum. We verify this numerically. For $\omega = $10.0 MeV, the ground-state CM energy $\mathcal{E}_0 = 7.47$ MeV is produced by the underlying $NN$ interaction, and it is taken as the reference energy: $\Eref = \mathcal{E}_0$. Tuning $C^{(0)}(\Eref; \Lambda)$ in the LO EDT potential \eqref{eqn:V1s0LO} for any given $\Lambda$ to reproduce $\mathcal{E}_0$ as the ground-state energy, we can then calculate the phase shifts in the neighborhood of the CM energy $E = \Eref$. If we restrict ourselves to a statement on the phase shift for exactly $E = \Eref$, and nowhere else, we expect to rediscover the same result predicted by the BERW formula. This is indeed the case, as shown in Fig.~\ref{fig_busch_lo_1s0}.

\begin{figure}[tb]
  \centering
  \includegraphics[scale=0.45]{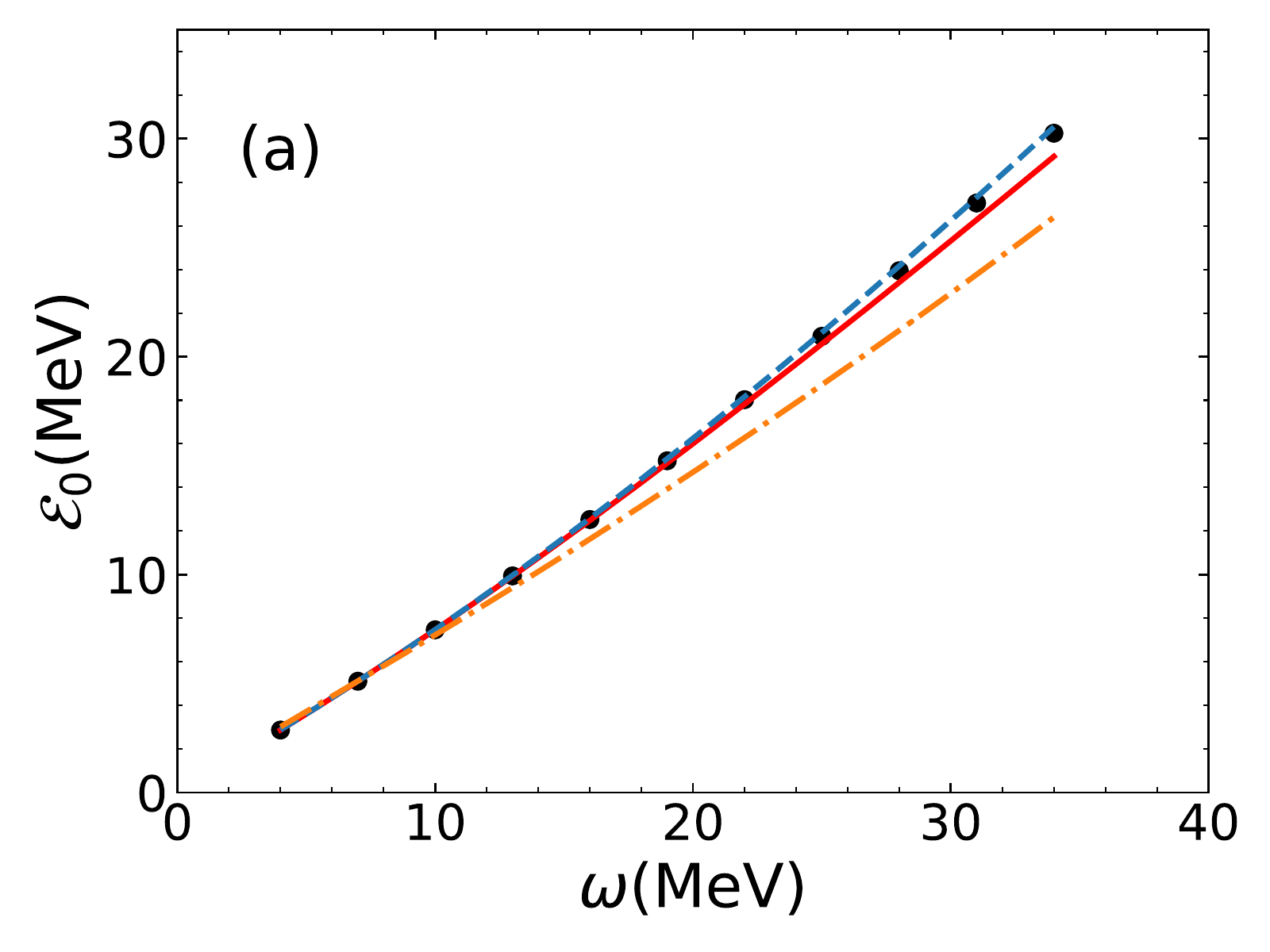}
  \includegraphics[scale=0.45]{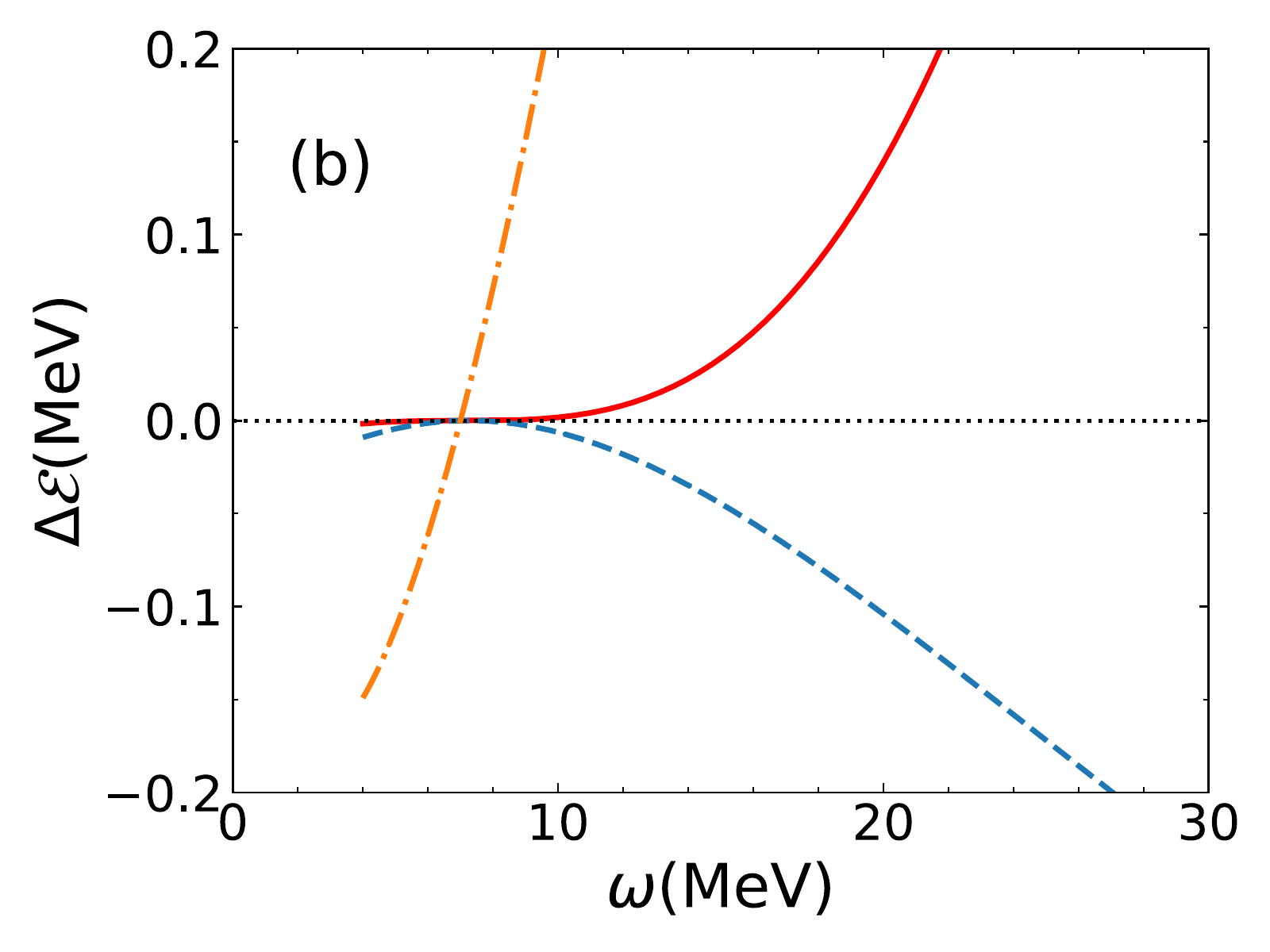}
  \caption{(a) $\cs{1}{0}$ ground-state energies as function of HO frequency $\omega$. The circles represent the data, the dot-dashed line LO of the EDT, the dashed line NLO, and the solid line {\NNLO}. (b) Similar to (a), but the deviation from the data, energies generated by the underlying $NN$ potential, at each order is shown instead.}
  \label{fig_1s0_energy_vs_omega}
\end{figure}

Let us turn to higher orders and to investigate what kind of improvement can be gained beyond the BERW formula. Before looking at scattering processes, we examine the spectra calculated with the higher-order EDT potentials. The ground-state energy as a function of $\omega$ is shown in Fig.~\ref{fig_1s0_energy_vs_omega} (a). The data at three different frequencies, $(\omega - \Delta \omega, \omega, \omega + \Delta \omega) = (6.5, 7.0, 7.5)$ MeV, are used as the inputs to determine the LECs. The determination is arranged so that the LO EDT curve goes through the datum at $\omega = 7.0$ MeV, the NLO through both $7.0$ and $7.5$ MeV, and finally the {\NNLO} through all of the data. Unless noted otherwise, when fitting the EDT LECs,  we will always pick the  data in the same fashion: usually three of them, one for each of three values of $\omega$. The predictions by the EDT for energy eigenvalues at higher frequencies are systematically improved with increasing orders. To better visualize the improvement, the deviation from the data $\Delta \mathcal{E}$ of each order is plotted in Fig.~\ref{fig_1s0_energy_vs_omega} (b).

If $\Delta \omega$ is quite small, the energies for $\omega \pm \Delta \omega$ can be computed from the eigen state for $\omega$ alone,
by considering small change of $\omega$ in first-order perturbation theory:
\begin{align}
    V_{\pm \Delta \omega} &\equiv \pm \mu \Delta \omega\, \omega r^2 \, , \label{eqn:VpmDelta_omega} \\
    \Delta \mathcal{E}_\pm &= \langle \Psi_\omega | V_{\pm \Delta \omega} | \Psi_\omega \rangle \, ,
\end{align}
where $\Psi_\omega$ is the eigen state for $\omega$. In other words, at least some of the LECs for a given $\Eref$ can be determined in practice with a fixed value of $\omega$, as opposed to varying $\omega$ for multiple times. This realization could be useful in larger-scale calculations where varying $\omega$ can be computationally expensive but the wave function of $\Psi_\omega$ can be saved in storage.

\begin{figure}[tb]
  \centering
  \includegraphics[scale=0.5]{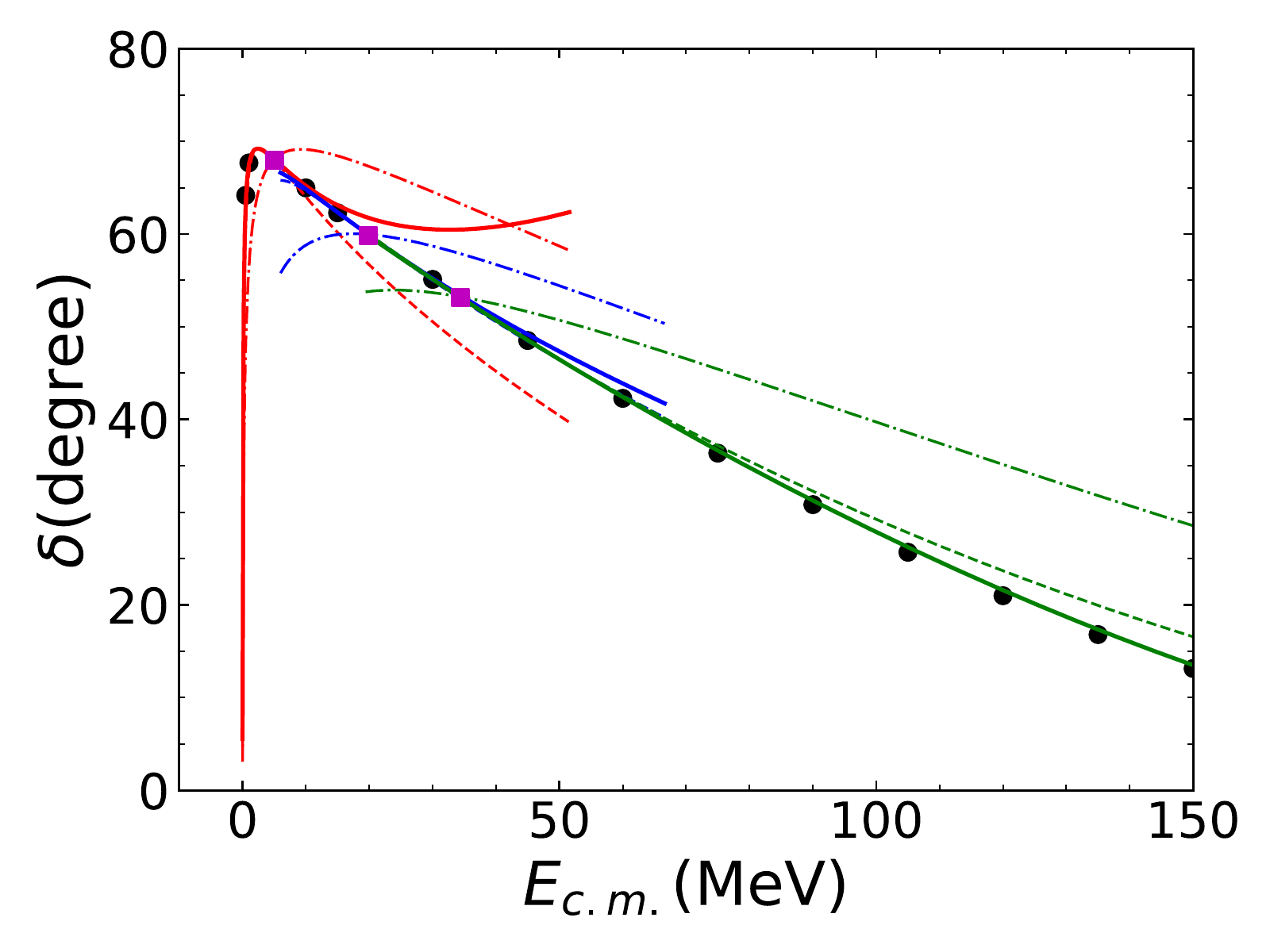}
  \caption{The $\cs{1}{0}$ phase shifts as a function of the CM energy. The circles represent those generated by the underlying $NN$ potential $V^\text{LO}_\chi$, the dot-dashed line is the LO of the EDT, the dashed line NLO, and the solid line {\NNLO}. The squares are the BERW values calculated with these three energy eigenvalues. There are three energy windows, corresponding to the ground, first, and second excited states for $\omega = 7$ MeV used as the inputs. In addition to the obvious sequence in energies, they are also distinguished by different colors.
}
  \label{fig_1s0_overall}
\end{figure}

The LECs obtained from fitting to the ground-state energy (Fig.~\ref{fig_1s0_energy_vs_omega}) are then used to predict the $\cs{1}{0}$ phase shifts. The same procedure is repeated with the first and second excited states, and $\Eref$ is adjusted accordingly so that it is always in close proximity to the energy eigenvalue under consideration. The results are plotted in Fig.~\ref{fig_1s0_overall}. Although it is quite encouraging that three states provide enough inputs to produce a good agreement with the chiral phase shifts, we do not have any a priori reason to expect an equally efficient result in other partial waves (cf. Sec.~\ref{sec:3p0}). But the EFT framework enables us to gauge how reliable the prediction of phase shifts is, by examining how rapid the results converge with increasing orders. We can find out the validity window for each member EFT identified with $\Eref$, by looking at where the EDT expansion diverges.

\section{P waves\label{sec:3p0}}

For the $S$ waves, the LO EDT potential is iterated to all orders; that is, the Lippmann-Schwinger equation is solved exactly in both scattering and trapping problems. But for higher partial waves, no immediate extension of such a nonperturbative iteration of contact potentials is known to be renormalizable (see Ref.~\cite{Epelbaum:2021sns} for the latest study), unless a dimeron field is explicitly used~\cite{Bertulani:2002sz}.

Let us briefly recapitulate the renormalization issue with the nonperturbatative treatment of purely short-range $P$-wave potentials, which has the following generic form after partial-wave projection:
\begin{equation}
    V_\text{P}(p', p; \Eref) = p^\prime p \left[C_\text{P}(\Eref) + \frac{1}{2}D_\text{P}({p'}^2 + p^2) + \frac{1}{2}E_\text{P}(\Eref)p'^2\,p^2 \cdots \right] \, . \label{eqn:PwavePot}
\end{equation}
The on-shell $\cp{3}{0}$ $T$-matrix obtained by iterating the $C_P$ term is
\begin{equation}
T(E) = -\frac{4\pi}{m_N} \frac{k^2}{\alpha_{P0} (\Eref) - \theta_1^\prime \Lambda (E - \Eref) - ik^3} \, ,
\end{equation}
where
\begin{align}
\theta_1^\prime &\equiv 4\pi m_N \theta_1 \, , \\
\alpha_{P0}(\Eref) &= -\frac{4\pi}{m_N} \left[C_\text{P}^{-1}(\Eref; \Lambda) - \theta_3 m_N \Lambda^3 - \theta_1 m_N^2 \Eref \Lambda \right]\, .
\end{align}
If we are interested in the phase shift at precisely $E = \Eref$, which is the case for the BERW formula, there will not be significant sensitivity to $\Lambda$ because $1/C^{(0)}_\text{p}(\Eref)$ can absorb the cubic divergence and the linear piece proportional to $\Eref$. But any prediction away from $E = \Eref$ is linearly dependent on $\Lambda$. So, with a straightforward nonperturbative treatment of the $C_\text{P}$ term, while we still recover the BERW formula for $P$ waves, the energy expansion we previously sought is lost.

There are two solutions. One is to resort to a power counting facilitated by the dimeron field, which is particularly useful if there are resonances in the $P$ waves~\cite{Bertulani:2002sz, Bedaque:2003wa}. The other is to exploit, if it is the case, the smallness of the $P$-wave phase shifts by developing a perturbative power counting. In fact, we can follow pionless EFT~\cite{Bertulani:2002sz} and let $C_P$, $D_P$, and $E_P$ parametrize each nonvanishing term of expansion in $(E - \Eref)$. To be more specific,
\begin{align}
    T^{(0)} & = V_P^{(0)} \, ,\\
    T^{(1)} & = V_P^{(1)} \, , \\
    T^{(2)} & = V_P^{(2)} + V_P^{(1)} G_0 V_P^{(1)} \, , \\
    T^{(3)} & = V_P^{(3)} + 2 V_P^{(2)} G_0 V_P^{(1)} + V_P^{(1)} G_0 V_P^{(1)} G_0 V_P^{(1)} \, ,
\end{align}
where
\begin{align}
    V_P^{(0)} &= 0 \, ,\\
    V_P^{(1)} &= C_P^{(0)} p' p \, , \\
    V_P^{(2)} &= p' p\left( C_P^{(1)}  + \frac{1}{2}D_P^{(0)}(p^2 + p^{\prime 2}) \right) \, , \\
    V_P^{(3)} &= p' p\left( C_P^{(2)}  + \frac{1}{2}D_P^{(1)}(p^2 + p^{\prime 2}) + \frac{1}{2} E_\text{P}^{(0)} p^2 p^{\prime 2}  \right) \, .
\end{align}
Again, the computation of energy levels of the trapped particles can be done by adding the HO potential.

\begin{figure}
  \centering
\includegraphics[scale=0.5]{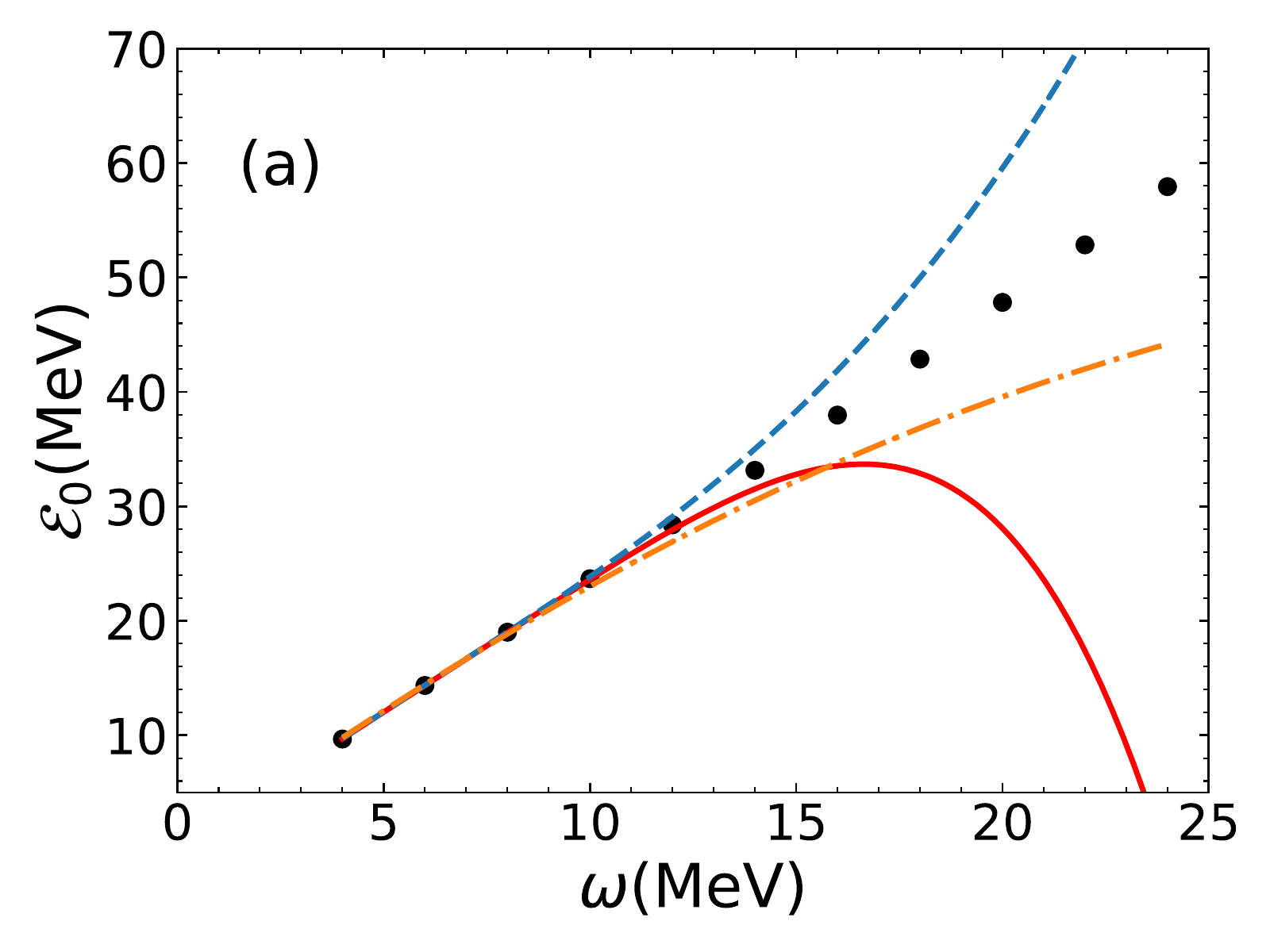}
\includegraphics[scale=0.5]{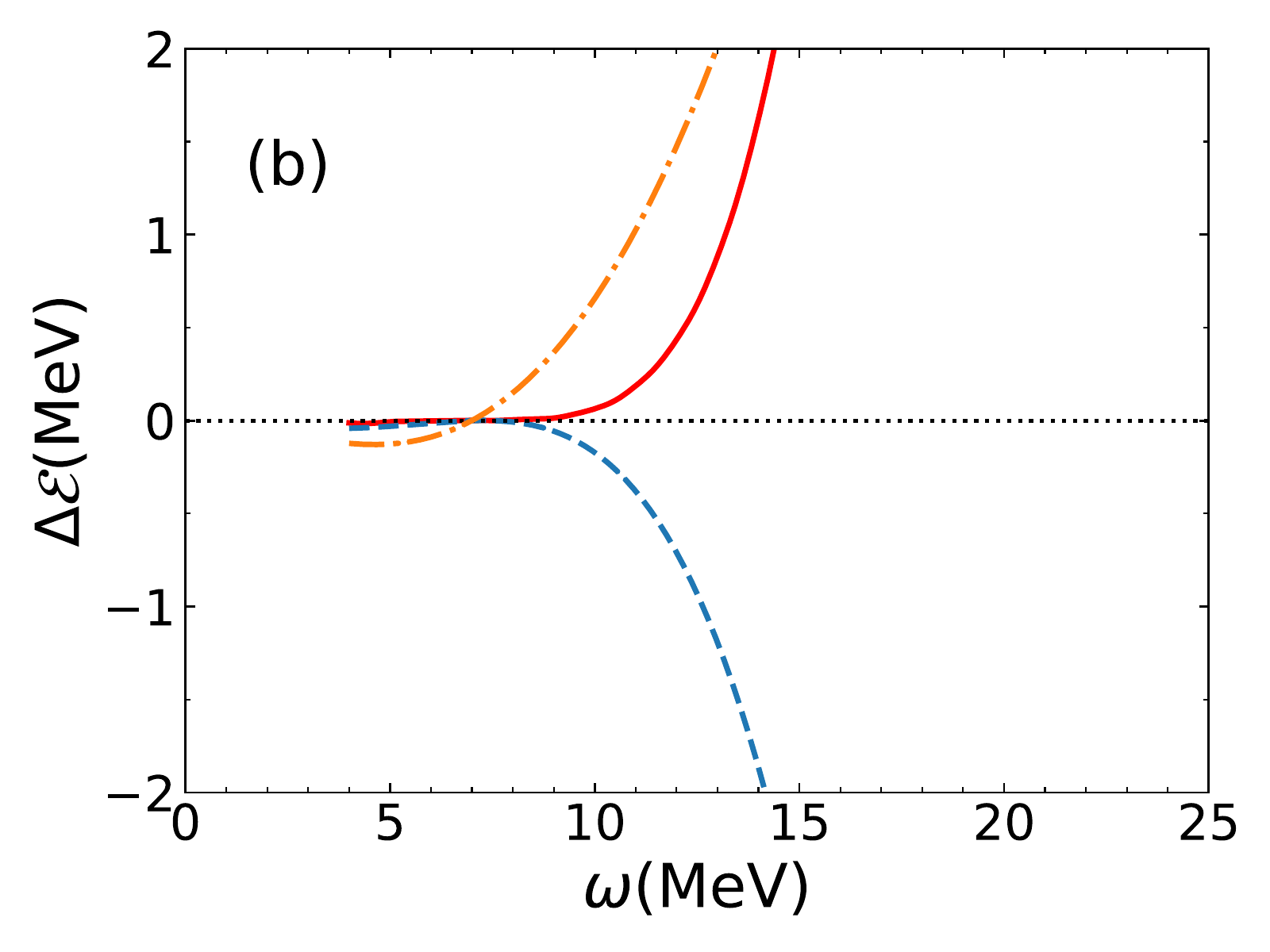}
  \caption{(a) The $\cp{3}{0}$ ground-state energy as a function of $\omega$. (b) The deviation from the data as a function of $\omega$. The symbols are the same as those of Fig.~\ref{fig_1s0_energy_vs_omega}.}
\label{fig:3p0_pert_levels}
\end{figure}

\begin{figure}
  \centering
\includegraphics[scale=0.5]{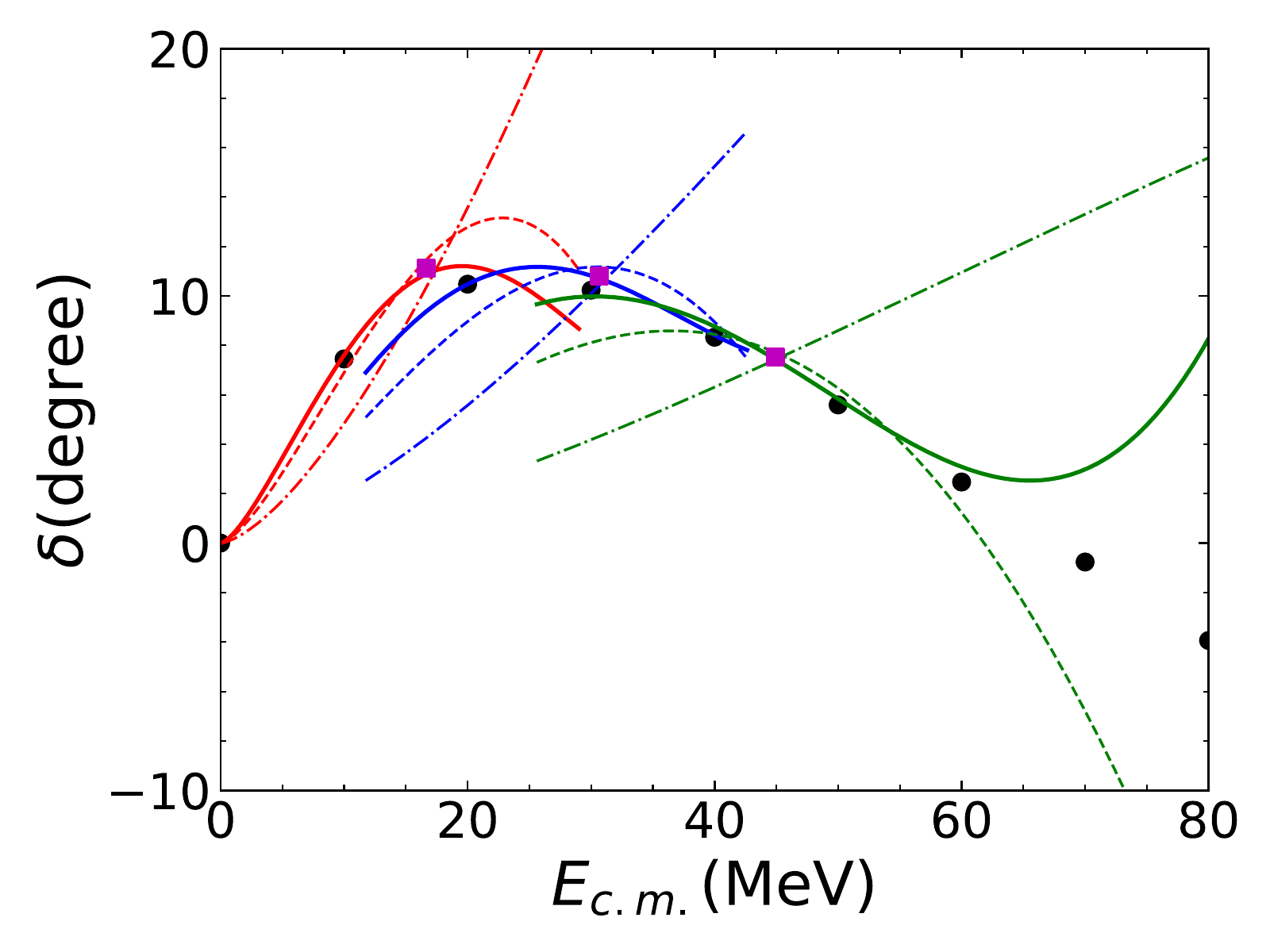}
  \caption{
  The $\cp{3}{0}$ phase shifts as a function of the CM energy. The lowest three states for $\omega = 7$ MeV are used as the inputs. The symbols are the same as those of Fig.~\ref{fig_1s0_overall}.
}
  \label{fig:3p0_pert_phase}
\end{figure}

The ground-state energy for the $\cp{3}{0}$ channel is shown in Fig.~\ref{fig:3p0_pert_levels}. It appears that the expansion converges slower than the $\cs{1}{0}$, which is also reflected by the phase shifts plotted in Fig.~\ref{fig:3p0_pert_phase} where the inputs from the first and second excited states are taken as well.

\section{$\omega$ dependence of scattering observables\label{sec:omega}}

As discussed at the beginning of Sec.~\ref{sec:framework}, model-independent extraction of scattering parameters from the trapped energy levels hinges on matching the scattering asymptotic wave function and the ``inside'' part of the trapped wave function in a non-interacting region where both trapping and intrinsic forces vanish. This requirement will not be met perfectly in the case of the HO trap, as long as a finite value is taken by $\omega$. Therefore, the scattering amplitudes computed in the previous sections under the assumption $\omega \to 0$ depend artificially on $\omega$, $T(E; \omega^2)$, a contamination related to the infrared scale $b$. In this section we discuss how to remove this artifact.

We expect to remove the implicit $\omega$ dependence of scattering amplitudes by subtracting the trapping force from within the range of the intrinsic interaction, which is equivalent to adding the opposite of the HO potential but only at short distances:
\begin{equation}
    V_\text{sub}(r; \omega^2) =  - \frac{1}{2} \mu \omega^2 r^2\, \theta (r_I - r) \, .
\end{equation}
This is in fact a change to the full Hamiltonian $H_\omega$, so the energy eigenvalues will shift from before the alteration of the HO trap. Consequently, one can no longer apply the BERW formula to the shifted energy levels. But we can always follow the recipe illustrated in the previous sections, calculating the energy levels with the altered trapping field, feeding them to the EDT potential encompassed by the same altered trapping field, and computing the phase shifts in the end. The end result will be the infinite-volume amplitude free of $\omega$: $T_\infty(E)$. If $\omega$ is sufficiently small, the calculations leading up to $T_\infty(E)$ from $T(E; \omega)$ can be done by treating $V_\text{sub}$ as a perturbation, for both energy eigenvalues and scattering amplitudes. This observation tells us that the discrepancy between $T(E; \omega)$ and $T_\infty(E)$ is a polynomial in $\omega^2$ for sufficiently small $\omega^2$:
\begin{equation}
    T(E; \omega^2) - T_\infty(E) =  c_1 \omega^2 + c_2 \omega^4 + \cdots \, . \label{eqn:T_omega_exp}
\end{equation}

\begin{figure}[tb]
  \centering
  \includegraphics[scale=0.45]{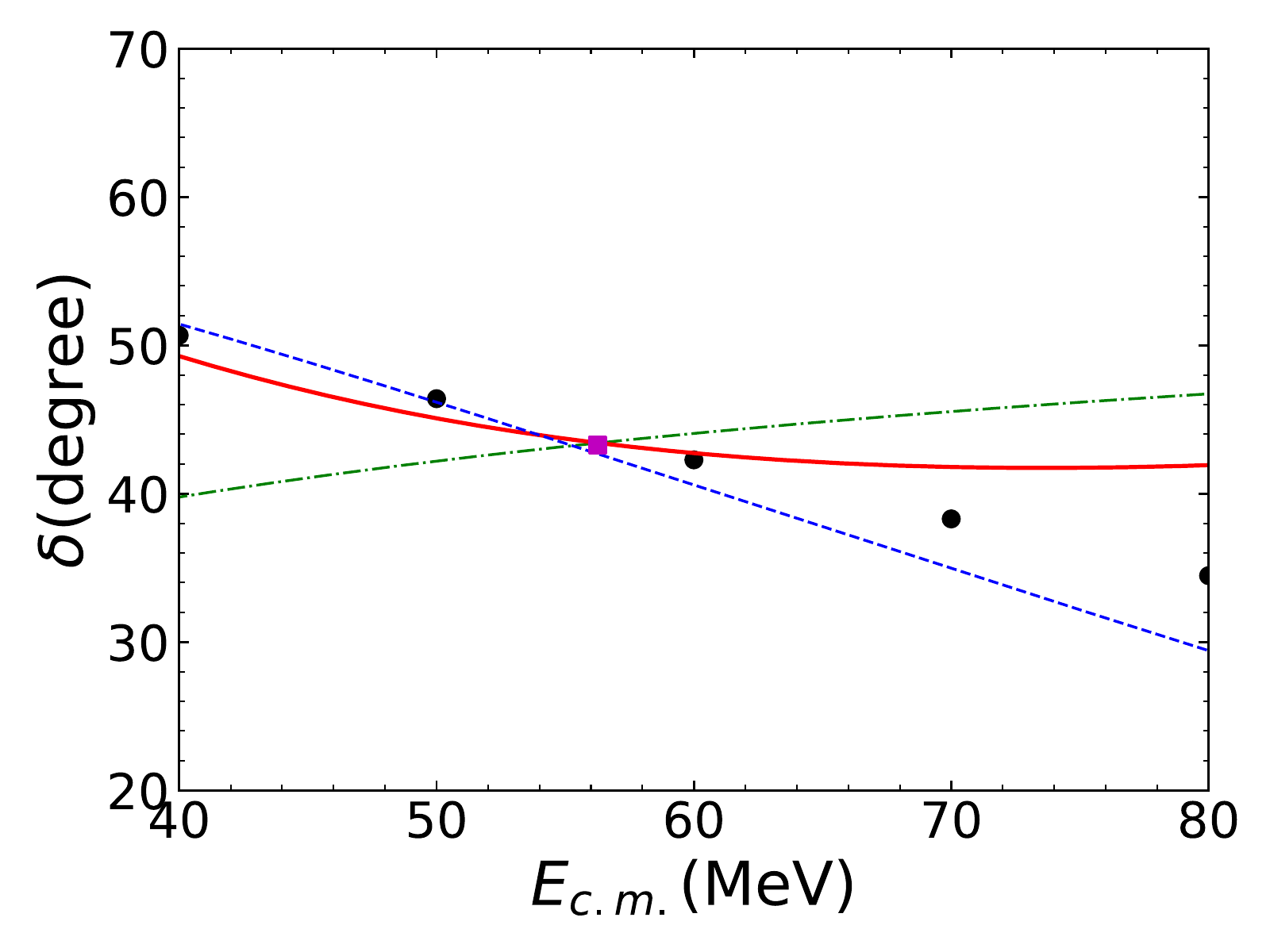}
  \caption{The $\cs{1}{0}$ phase shifts as a function of the CM energy, with the ground state for $\omega = 56$ MeV used as the input. See Fig.~\ref{fig_1s0_overall} for explanation of the symbols.
}
  \label{fig_1s0_phase_highenergies}
\end{figure}

If $r_I/b \ll 1$, which translates into
\begin{equation}
    \mu \omega r_I^2 \ll 1 \, ,
\end{equation}
there is an almost noninteracting region, so $\omega$ is expected to be within the convergence radius of Eq.~\eqref{eqn:T_omega_exp}. Examining the first-order correction,
\begin{equation}
    -\frac{1}{2} \mu \omega^2 \int^{r_I} dr r^4 \psi^2(r; E) \, , \label{eqn:omega_rI_int}
\end{equation}
we realize, however, that other scales embedded in the wave function could facilitate the perturbation theory. Therefore, the criterion $\mu \omega r_I^2 \ll 1$ could be unnecessarily conservative. In Fig.~\ref{fig_1s0_phase_highenergies}, the ground state for $\omega = 56$ MeV is used and the phase shifts close to $\Eref$ are only a few degrees off. Given that $\mu \omega r_I^2 \simeq 1.3$, where $r_I = m_\pi^{-1} \simeq 1.4$ fm, the surprisingly good agreement with the data can only be ascribed to the short-range structure of the wave function $\psi(r; E)$. Analyzing the wave function can be difficult, especially in ab initio calculations for many-nucleon systems. So we refrain from making a general statement about the precise convergence radius of the series \eqref{eqn:T_omega_exp}.

The EDT framework does not automatically address the issue of removing the infrared artifacts, but we are better equipped to tackle it. With the BERW formula alone, one can not vary $\omega$ while fixing the value of the CM energy $E$ at which the amplitude is calculated. This can be done now in the EDT framework. We note that Ref.~\cite{Zhang:2019cai} used a different strategy, building into the EFT Lagrangian interaction terms proportional to $\omega^2$.

As an application, we show how to extract the $\cs{1}{0}$ scattering length $a_{\cs{1}{0}}$, which is an observable associated with $E = 0$. Tabulated in Table~\ref{tab:a1s0_delta_wdot5} are the values of $a_{\cs{1}{0}}$ calculated for various $\omega$'s, up to {\NNLO} in the EDT expansion:
\begin{equation}
a_{\cs{1}{0}} = a^{(0)} + a^{(1)} + a^{(2)} + \cdots
\end{equation}
Besides extrapolating to the threshold, the EDT offers a means to assess truncation uncertainty of higher order in $(E - \Eref)$. We estimate $a^{(3)}$ to be order of
\begin{equation}
   \sim \Bigg{|} a^{(2)} \frac{a^{(2)}}{a^{(1)}} \Bigg{|} \, ,
\end{equation}
which is used as the EDT truncation uncertainty and listed in the column with header $\Delta a$ in Table~\ref{tab:a1s0_delta_wdot5}.

\begin{table}
  \centering
\begin{tabular}{|c|c|c|c|c|}
  \hline
 \ $\omega$ (MeV) & \ $a^{(0)}$\ & \ $a^{(1)}$\ & \ $a^{(2)}$\ & $\Delta a$ \\
  \hline
  6.0 & -9.66  &-5.95   &-3.24   &1.8    \\
  \hline
  8.0 & -7.69   &-5.66   &-3.45   &2.1    \\
  \hline
  10.0 & -6.59  &-5.34   &-3.45   &2.2    \\
  \hline
  12.0 & -5.74  &-5.02   &-3.40   &2.3    \\
  \hline
\end{tabular}
\caption{$a_\cs{1}{0}$ (fm) calculated at each order with various values of $\omega$. $\Delta a$ is the EDT truncation uncertainty. \label{tab:a1s0_delta_wdot5}}
\end{table}

\begin{figure}
  \centering
\includegraphics[scale=0.5]{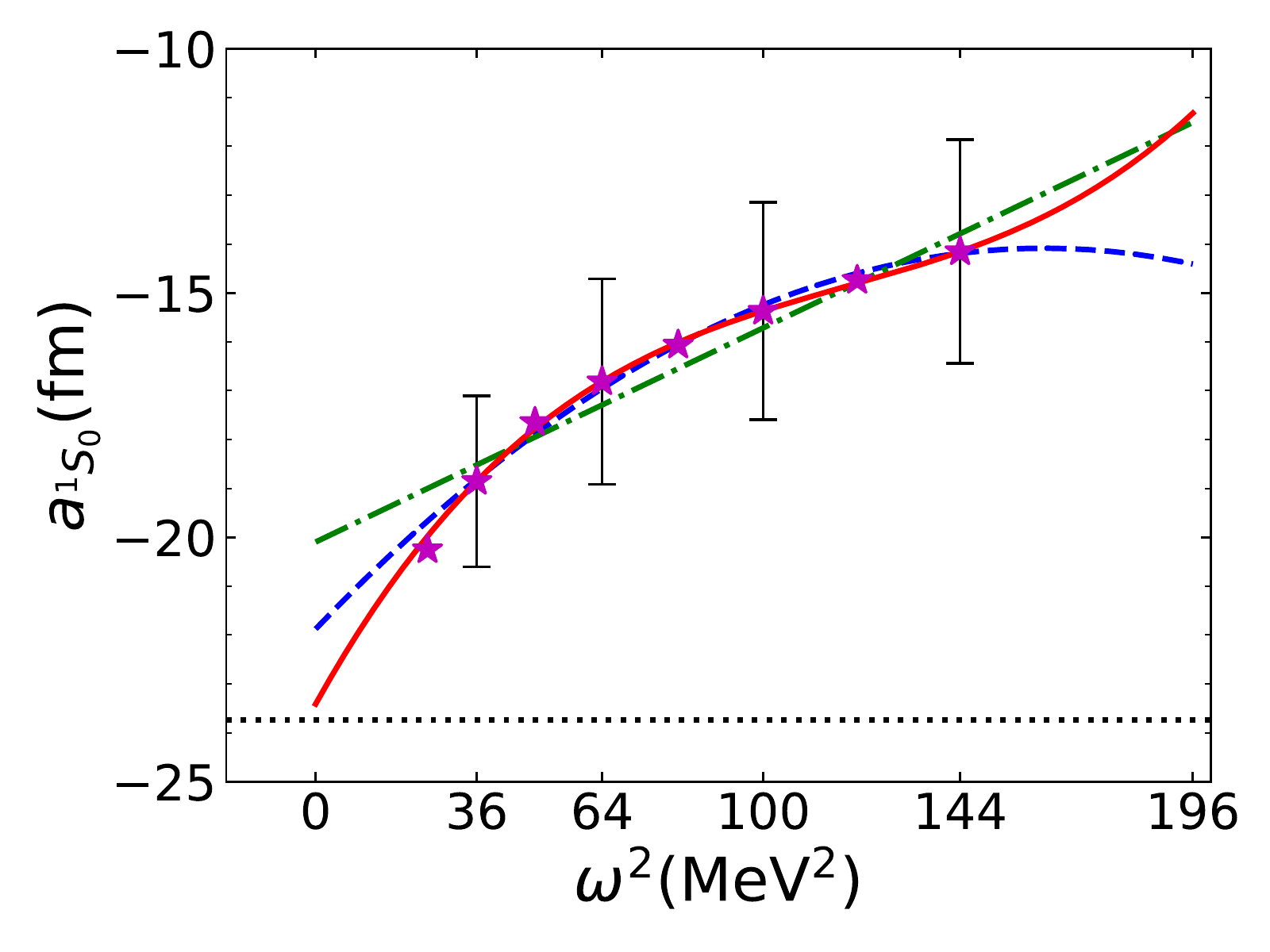}
  \caption{Extrapolation of $a_\cs{1}{0}(\omega)$ to the infinite-volume limit. The stars represent the {\NNLO} EDT result of $a_\cs{1}{0}$ for different values of $\omega^2$. The dot-dashed, dashed, and solid lines are respectively linear, quadratic, and cubical polynomials in $\omega^2$, fitted to four inputs of $a_\cs{1}{0}(\omega)$ that have their EDT truncation uncertainty $\Delta a$ shown by the bars.
}
  \label{fig:a1s0_wdependence_withw0error}
\end{figure}

Following Eq.~\eqref{eqn:T_omega_exp}, we expect $a_\cs{1}{0}(\omega)$ to be a polynomial in $\omega^2$:
\begin{equation}
    a_{\cs{1}{0}}(\omega) = a_\infty + c_1 \omega^2 + c_2\omega^4 + c_3 \omega^6 + \cdots
\end{equation}
To extract the infinite-volume limit of the scattering length $a_\infty$, we fit the above polynomial to the values of $a^{(2)}$ in Table~\ref{tab:a1s0_delta_wdot5}, using a least square weighted by the truncation uncertainty $\Delta a$. The fits with increasing polynomial degree are performed, and the results are plotted in Fig.~\ref{fig:a1s0_wdependence_withw0error}. $a_\cs{1}{0}(\omega)$ for some other values of $\omega$ are also shown even though they were not employed in the fits. With increasing degree in $\omega^2$, the value of $a_\infty$ approaches its ``true'' value of $-23.74$ fm, with $-23.4$ fm resulting from the fit to the third-degree polynomial.

\section{Discussions and conclusions\label{sec:conclusions}}

We have proposed a framework of energy-dependent effective field theory to convert the energy levels of trapped particles to their scattering amplitudes. The EDT developed for this purpose is a collection of contact EFTs. Unlike the more common hierarchy of EFTs where the underlying one also applies to the lower-energy region, each member EFT of the EDT specializes in describing a certain kinematic region marked by a reference kinematic parameter. In the case of two-particle elastic scattering, this reference kinematic parameter is chosen to be the CM energy. The Lagrangians of these member EFTs have the same set of interaction terms, but with LECs depending on the reference energy:
\begin{equation}
    \mathcal{L}_\text{EDT} = \sum_{\Eref} \mathcal{L} \left[C_i({\Eref})\right] \, .
\end{equation}
Each member EFT underpins an expansion of scattering amplitudes and trapped states around its reference energy; therefore, it can connect observables from both sides. It will not be surprising if the EDT does not offer strong predictive power, but this is hardly a concern because its usefulness lies in extracting scattering observables from energy levels; how much is gained relies on how much is invested.

The BERW formula is precisely the LO of the said expansion, and one can now reliably predict the phase shifts at energies different from the trapped states, thanks to the systematic approximation provided by the EDT framework. One might be discouraged by the increasing number of LECs required by higher orders because it demands more inputs from ab initio calculations of the energy levels. We have argued, however, that more inputs of energy eigenvalues for varied $\omega$ can be obtained by first-order perturbation theory, which entails only the wave function for $\Eref$. This has the potential to save computational costs, as opposed to explicitly varying $\omega$ in ab initio calculations.

In the illustrative application of $\cs{1}{0}$ $NN$ scattering, we have based the power counting on expansion of the $T$ matrix around the reference energy [see Eq.~\eqref{eqn:T1s0_exp}]. But it is quite likely that we need to develop distinctive power counting for different systems. For example, near a resonance, a two-parameter interaction is needed to set up the LO so that both energy and width of the resonance can be captured. Another example is demonstrated in Sec.~\ref{sec:3p0} by the application to the $P$ waves, using a perturbative power counting by exploiting smallness of the $\cp{3}{0}$ $NN$ phase shifts.

The presence of the trapping force within the range of intrinsic interactions leads to dependence on the frequency $\omega$, an unwanted infrared artifact. We showed that the artifact is a polynomial in $\omega^2$ and can be removed by extrapolating to the limit $\omega^2 = 0$.

For the particular problem of two-particle elastic scattering, using the full EDT machinery may seem an overkill. One could approximate the energy eigenvalue as a function of $\omega$ with a Taylor expansion:
\begin{equation}
    \mathcal{E}(\omega) = \mathcal{E}_0 + f_1 (\omega - \omega_0) + f_2(\omega - \omega_0)^2 + \cdots \, ,
\end{equation}
which, when coupled with the BERW formula, will also yield the phase shifts for a certain energy region. But the EDT's ability to extrapolate to the continuum limit is lost because the BERW formula does not allow one to calculate the amplitude at a fixed energy while varying $\omega$~\cite{Zhang:2019cai}.

The EDT framework offers encouraging prospect for applications to more complicated nuclear reactions. For a specific reaction, it is quite straightforward to generalize to a cluster/halo EFT Langrangian ~\cite{Hammer:2017tjm}, using the nuclear clusters participating in the reaction as the degrees of freedom, with the LECs depending on a reference energy, or other kinematic variables. On the side of ab initio calculations, energy levels of all the interacting constituent nucleons are produced and used to determine the LECs. A similar program was carried out to study elastic scattering of the neutron by a nucleus in Ref. ~\cite{Zhang:2020rhz}, but limited to the validity region of the conventional pionless or cluster EFT. We can now make use of the EDT framework laid out in the paper and go beyond that limitation. The Coulomb force ~\cite{Guo:2021qfu} and coupled-channel effects can be accounted for with relative ease ~\cite{Guo:2021uig}.

But there are more serious obstacles to overcome before artificial HO trapping becomes a general facility for ab initio description of nuclear reactions. For a given value of the CM energy of the reaction under consideration, what are the values of $\omega$ one needs to cover in the ab initio calculations in the HO trap so that a reliable infinite-volume extrapolation by way of $\omega^2$ polynomials ~\eqref{eqn:T_omega_exp} can be achieved? There should be a method in place to estimate these values beforehand. This is further complicated by the size d of each cluster if d is comparable with other scales of the system.

Yet another issue to be addressed is three-body or even higher-body channels in some reactions. For instance, for nucleon-deuteron breakup, the two-body framework must be extended to incorporate the three-body final states. Even if only elastic scattering cross section is of interest, the mixing is always present in the trap between two-body and three-body states above the breakup threshold, which cannot be dealt with in a pure two-body formalism.

\acknowledgments

We thank Xilin Zhang and Xu Feng for useful discussions. The work was supported in part by the National Natural Science Foundation of China (NSFC) under Grants No. 11775148 and No.11735003.

\bibliography{HOtrapEFT.bib}

\end{document}